\documentclass[11pt, english]{article}
\pdfoutput=1
\usepackage{fullpage}
\usepackage{amsmath}
\usepackage{amssymb}
\usepackage{amsthm}
\usepackage{pxfonts}
\usepackage[T1]{fontenc}
\usepackage{color}
\usepackage{graphicx}
\usepackage{float}
\usepackage{hyperref}
\usepackage{charter}

\newtheorem{theorem}{Theorem}
\newtheorem{lemma}[theorem]{Lemma}

\newtheorem{corollary}[theorem]{Corollary}

\newtheorem{openproblem}{Open Problem}

\def\a{{\alpha}}
\def\b{{\beta}}
\def\g{{\gamma}}
\newcommand{\lemlab}[1]{\label{lemma:#1}}

\newcommand{\figlab}[1]{\label{fig:#1}}

\newcommand{\lemref}[1]{\ref{lemma:#1}}

\newcommand{\figref}[1]{\ref{fig:#1}}

\definecolor{changes}{rgb}{0,0.2,0.6}

\title{Cauchy's Arm Lemma on a Growing Sphere}

\author{
  Zachary Abel%
    \thanks{Department of Mathematics, Harvard University,
      1 Oxford Street, Cambridge, MA 02138, USA.
      \protect\url{zabel@fas.harvard.edu}}
\and
  David Charlton%
    \thanks{Department of Computer Science, Boston University,
      111 Cummington Street, Boston, MA 02215, USA.
      \protect\url{charlton@cs.bu.edu}}
\and
Sébastien Collette\thanks{Chargé de recherches du F.R.S.-FNRS, \protect\url{sebastien.collette@ulb.ac.be}. Computer Science Department, Universit\'e Libre de Bruxelles, CP212, Bvd. du Triomphe, 1050 Brussels, Belgium}
\and
  Erik D. Demaine%
    \thanks{MIT Computer Science and Artificial Intelligence Laboratory,
      32 Vassar Street, Cambridge, MA 02139, USA,
      \protect\url{{edemaine,mdemaine}@mit.edu}}
    \thanks{Partially supported by NSF CAREER award CCF-0347776,
            DOE grant DE-FG02-04ER25647, and AFOSR grant FA9550-07-1-0538.}
\and
  Martin L. Demaine\footnotemark[4]
\and
Stefan Langerman\thanks{Chercheur qualifié du F.R.S.-FNRS, \protect\url{stefan.langerman@ulb.ac.be}. Computer Science Department, Universit\'e Libre de Bruxelles, CP212, Bvd. du Triomphe, 1050 Brussels, Belgium}
\and
  Joseph O'Rourke%
    \thanks{Department of Computer Science, Smith College,
        Northampton, MA 01063, USA, \protect\url{orourke@cs.smith.edu}}
\and
  Val Pinciu%
    \thanks{Department of Mathematics,
Southern Connecticut State University, USA, \protect\url{pinciu@scsu.ctstateu.edu}}
\and
  Godfried Toussaint%
   \thanks{School of Computer Science, McGill University, 3480 University St., Montreal, Canada, \protect\url{godfried@cs.mcgill.ca}}
}

\begin{document}
\maketitle
\sloppy
\begin{abstract}
We propose a variant of Cauchy's Lemma, proving that when a convex chain on one sphere is redrawn (with the same lengths and angles) on a larger sphere, the distance between its endpoints increases. The main focus of this work is a comparison of three alternate proofs, to show the links between Toponogov's Comparison Theorem, Legendre's Theorem and Cauchy's Arm Lemma. 
\end{abstract}

\section{Introduction}
A chain is composed of a sequence of points $v_{0}, \ldots, v_{n}$, where each consecutive pair is connected by edges $e_{i}=v_{i-1}v_{i}$, and the angle at vertex $v_{i}$ between $e_i$ and $e_{i+1}$ is $\theta_{i}$. A chain is convex if the corresponding polygon (that is, if we add the edge $v_{0}v_{n}$ to close the chain) is convex. These are well defined in the plane as well as on the sphere. In the case of the sphere, the edges are arcs of great circles. Cauchy's lemma can be expressed in the following way (although this was not its original statement):
\begin{lemma}[Cauchy]\label{lem:cauchy}
Given a convex chain $\mathcal C$, if we increase the value of some nonempty subset of the angles $\theta_{i}$, while keeping the length of the edges fixed and every $\theta_{i}\le \pi$, then the distance between the endpoints $v_{0}$ and $v_{n}$ strictly increases.
\end{lemma}
\noindent Although the original proof was flawed, this lemma was later proved correctly in the plane, as well as when the chain is on the surface of a sphere~\cite[p.~228]{Cromwell}.

In this note, we consider the following question: what happens to the endpoints of a chain on a sphere if, instead of choosing some angles to increase, we change the \emph{radius} of the sphere while preserving both the edge lengths and the angles?  

Spherical geometry tells us that when convex polygons are drawn on the sphere, the sum of their angles is larger than in the convex polygons with same edge lengths in the plane. This increment is called the \emph{spherical excess}. Its value decreases when the radius of the sphere grows, and also depends on the area of the polygon. Intuitively, as the sum of the angles decreases when the sphere gets larger, one might think that, given a convex chain $\mathcal C$ on the sphere $\mathcal S$ and a chain with same angles and edge lengths $\mathcal C'$ on a larger sphere $\mathcal S'$, the endpoints of $\mathcal C'$ should be farther from each other than those of $\mathcal C$. 

However, this intuition alone does not suffice to reach this conclusion, because even though the sum of the angles is larger in $\mathcal C$, it is not certain that every angle individually grows. This is what we show in the next sections, by analyzing the behavior of the angles of a triangle when drawn on different kinds of surfaces.\bigskip

Most results in this note follow from a theorem in Riemannian geometry, Toponogov's Comparison Theorem~\cite{toponogov}, which we discuss in Section~\ref{sec:toponogov}. Toponogov's version is strictly more general as it covers arbitrary manifolds with positive curvature, of which our "growing sphere" version is a special case. 

However, we believe that the comparison of these approaches is of general interest because (1)~the available proofs of Toponogov's Theorem rely on machinery in Riemannian geometry, whereas our proof is elementary; and (2)~the statement of Cauchy's Arm Lemma in the alternate form provided here might be useful in further research directions.
\color{black}

\section{Cauchy's Lemma from the Sphere to the Plane}
In this section, we compare a chain drawn on a sphere to a similar chain drawn in the plane. Note that the plane is equivalent to a sphere of infinite radius, so this is clearly a subproblem of the original question. 
Our result is formally stated as follows: 

\begin{theorem}\label{thm:spheretoplane}
Let $\mathcal C$ be a convex chain embedded on the sphere $\mathcal S$, and $\mathcal C'$ be a convex chain with same angles and edge lengths in the plane. 
Then the distance between the endpoints of $\mathcal C'$ is greater than the distance between the endpoints of $\mathcal C$.
\end{theorem}

To derive this result, we first prove that given a triangle on the sphere $\mathcal S$ and a triangle with the same edge lengths in the plane, the latter has smaller angles. A more general proof of this will be provided in the next section. We believe, however, that this proof is interesting as it involves neither Legendre's Theorem nor Toponogov's Comparison Theorem. 

\begin{lemma}
Let $\triangle abc$ be a geodesic triangle on a sphere,
with side lengths $A$, $B$, and $C$ opposite vertices
$a$, $b$, and $c$ respectively.
Let $b'$ and $c'$ be the midpoints of $ab$ and $ac$
respectively.  Then the length $A'=|b'c'|$
of this ``midchord''
is strictly greater than $A/2$.
\lemlab{Val.lemma}
\end{lemma}

\noindent
Note that, in the plane, we would have $A' = A/2$.

\noindent
\begin{proof}
Let the given triangle have angles $\a$, $\b$, $\g$
at vertices $a$, $b'$, $c'$ respectively.
Extend the geodesic segment $b'c'$ to $b''$ so that $|c'b''|=A'$,
and connect $b''$ to $c$.
See Figure~\figref{spherical_triangles_Val}.
\begin{figure}[htbp]
\centering
\includegraphics[width=0.75\linewidth]{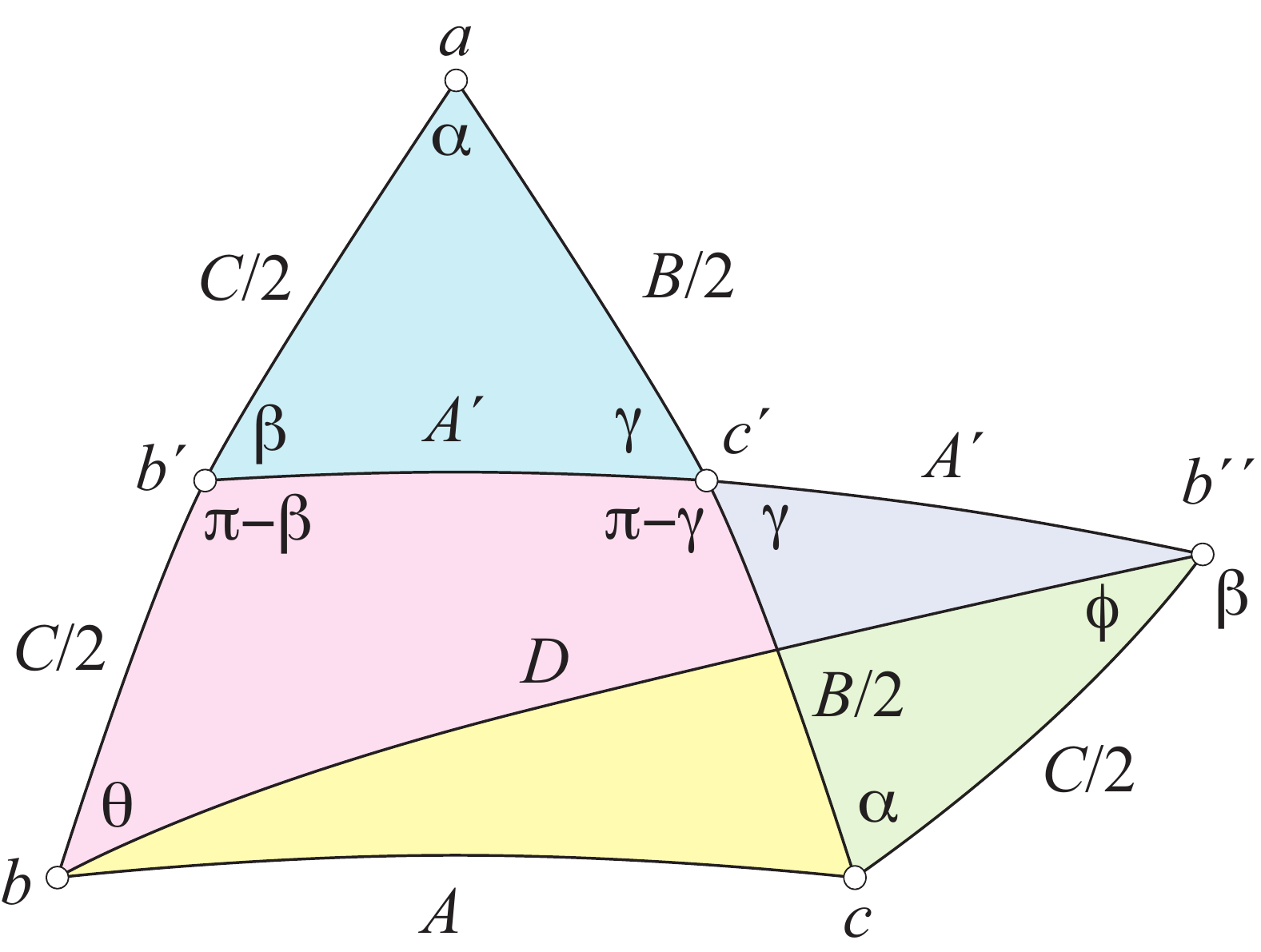}
\caption{$\triangle c b'' c'$ is congruent to $\triangle a b' c'$,
and $\theta > \phi$.}
\figlab{spherical_triangles_Val}
\end{figure}
Then $\triangle c b'' c'$ is congruent to $\triangle a b' c'$,
as it has the same angle $\g$ at $c'$, included between
the same side lengths $A'$ and $B/2$.

Now draw the geodesic diagonal $b b''$, and call the angles
on its opposite sides $\theta$ and $\phi$
as illustrated.
For the upper triangle $\triangle b b' b''$, we
have 
$\theta + (\pi-\b) + (\b-\phi) > \pi$
(because the angles of a spherical triangle sum to strictly greater
than $\pi$),
i.e., $\theta > \phi$.
Now notice that the two sides of the upper triangle
determining $\theta$ have lengths $D$ and $C/2$,
and the two sides of the lower triangle
determining $\phi$ have the same lengths $D$ and $C/2$.
Therefore, $\theta > \phi$ implies the same inequality
in the lengths of the opposite sides: $2 A' > A$.
This establishes the claim of the lemma.
\end{proof}

\begin{lemma}
Let a planar triangle have side lengths $A$, $B$, and $C$.
Each angle of the (unique) spherical triangle with the same side lengths
is strictly larger than the corresponding planar angles.
\lemlab{X}
\end{lemma}
\begin{proof}
Let the planar triangle be
$P=\triangle abc$ with angle $\a$ at $a$,
and the spherical triangle be
$S=\triangle a'b'c'$ with angle $\a'$ at $a'$.
We prove that $\a' > \a$.

We draw midchords in both $P=P_0$ and $S=S_0$
as in Lemma~\lemref{Val.lemma},
bisecting the sides of length $B$ and $C$.
Call these midchords $p_1$ and $s_1$ in $P_0$ and $S_0$ respectively,
and call the triangles above these midchords
$P_1$ and $S_1$ respectively.
See Figure~\figref{exponen_seq_tri}.
We know that $|p_1| = A/2$ because $P$ is planar,
and that $|s_1| > A/2$ by Lemma~\lemref{Val.lemma}.
Repeat the construction on $P_1$ and $S_1$, with
midchords $p_2$ and $s_2$.
We have $|p_2| = |p_1|/2 = A/4$,
and $|s_2| > |s_1|/2 > A/4$.
Continuing in this manner, for any $i$,
we obtain $|p_i| = A/2^i$ and  $|s_i| > A/2^i$.

Let $B_i=B/2^i$ and $C_i=C/2^i$ be the side lengths of the triangles
$P_i$ and $S_i$ after $i$ iterations.
Note that these side lengths are the same by construction
for the planar and spherical triangles.
Applying the law of cosines to the planar triangle, we have
$$
\a = \arccos \left[ \frac{1}{2} \left( B_i/C_i + C_i/B_i - |p_i|^2/(B_i C_i) \right) \right]
$$
Note that the length ratios in this expression are independent of $i$,
as all the triangles $P_i$ are similar to the starting triangle $P$.
As $i \rightarrow \infty$, the spherical triangles $S_i$ approach
planarity in the limit, so we have
$$
\a' = \lim_{i \rightarrow \infty}
\arccos \left[ \frac{1}{2} \left( B_i/C_i + C_i/B_i - |s_i|^2/(B_i C_i) \right) \right]
$$
Because the ratios $B_i/C_i=B/C$ and $C_i/B_i=C/B$ are constant,
the only difference between the $\a$ and $\a'$ expressions
occurs in the $|p_i|$ and $|s_i|$ factors.
But we know that $|s_i| > |p_i|$ for all $i$, and indeed
$|s_i| - |p_i| > |s_1| - |p_1| > 0$, so
$$
\lim_{i \rightarrow \infty} |s_i|^2/(B_i C_i)
$$
is strictly greater than the constant
$|p_i|^2/(B_i C_i)=A^2/(B C)$.
Because the $\arccos$ function is monotonically decreasing,
this implies that $\a' > \a$, as claimed.

Repeating the argument for each angle of the triangle establishes
the lemma.
\end{proof}

\noindent
Thus, when a planar triangle is drawn on a sphere
(with the same side lengths), all angles
increase, and correspondingly, when a spherical triangle is
drawn in the plane, all angles decrease.

\begin{figure}[htbp]
\centering
\includegraphics[width=0.95\linewidth]{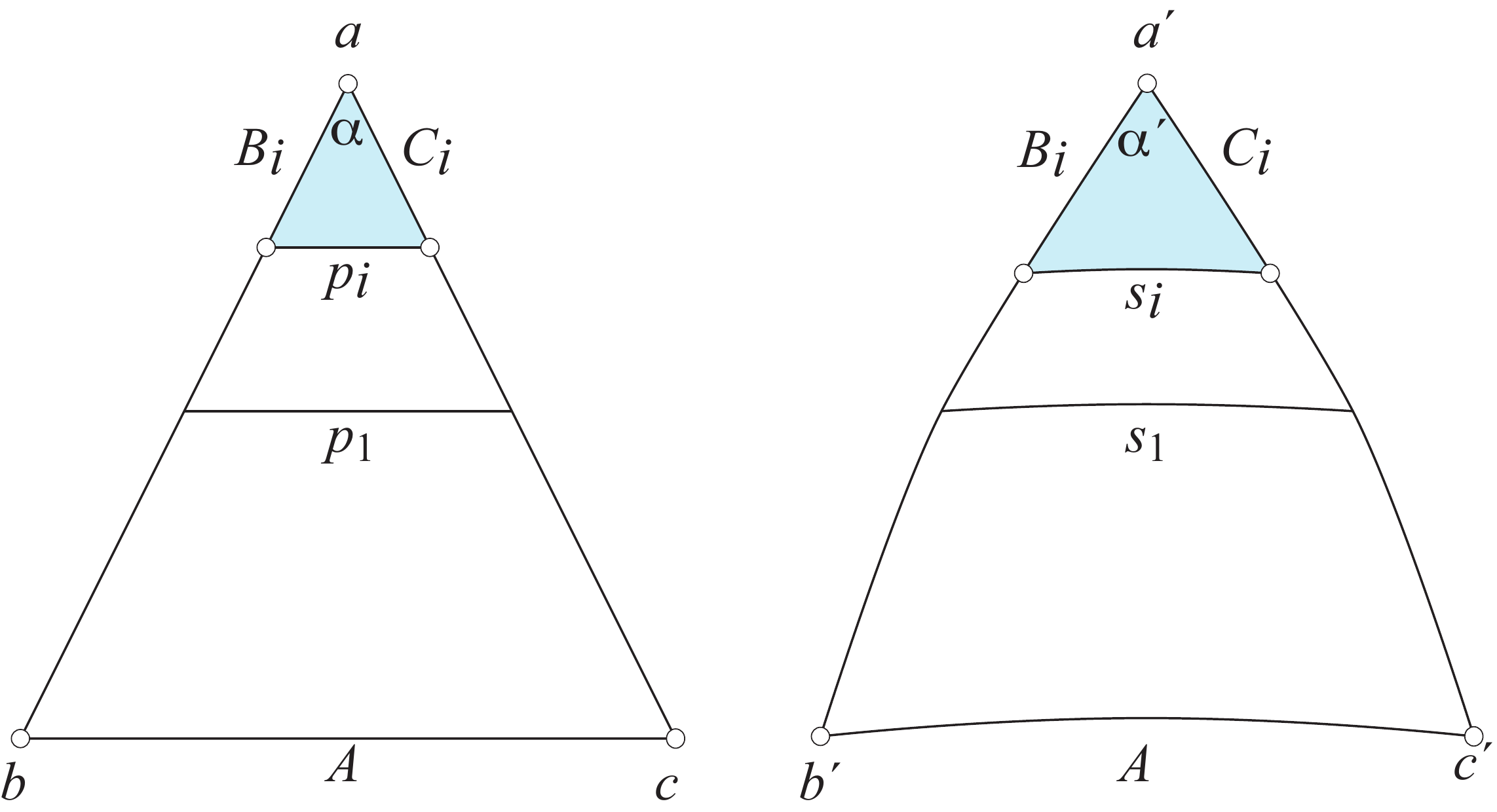}
\caption{$p_i$ is the $i$-th midchord in the planar triangle
(left), and $s_i$ in the spherical triangle (right).
$|s_i| > |p_i|$ and $\a' > \a$.
}
\figlab{exponen_seq_tri}
\end{figure}

\begin{figure}[htbp]
\centering
\includegraphics[width=0.95\linewidth]{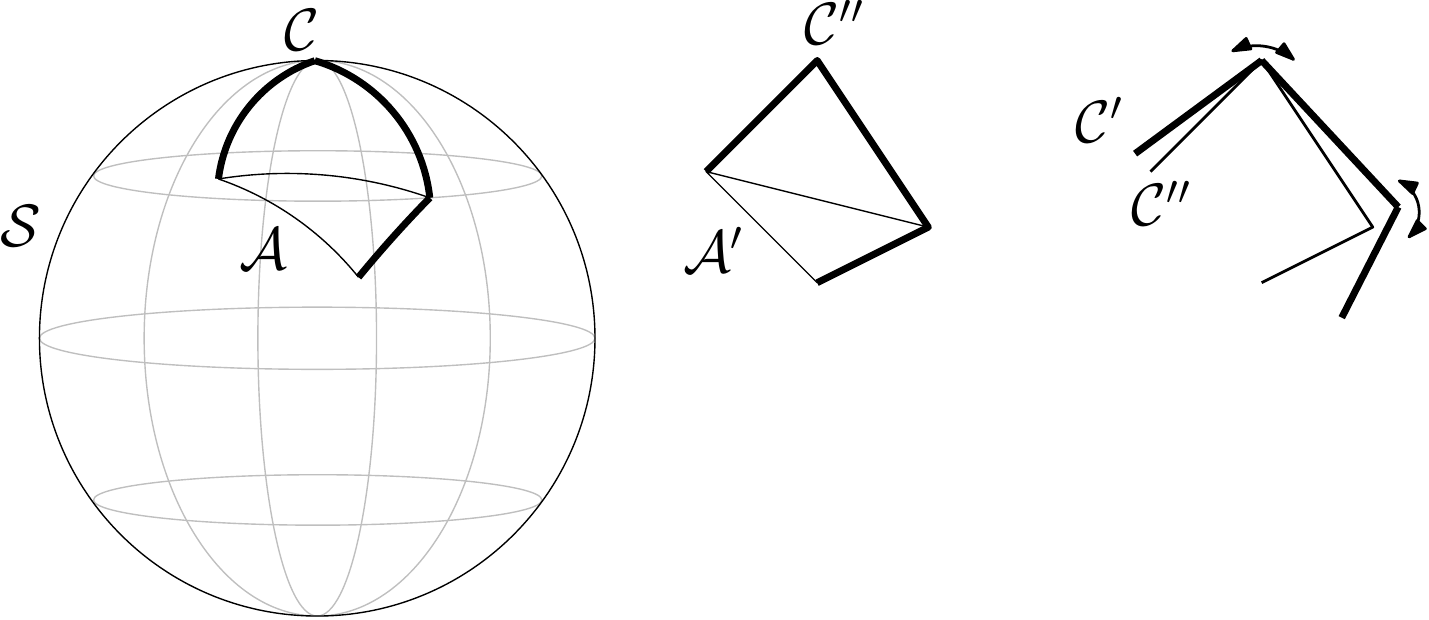}
\caption{Illustration of the proof of Theorem~\ref{thm:spheretoplane}.}
\label{fig:spheretoplane}
\end{figure}

\begin{proof}[Proof of Theorem~\ref{thm:spheretoplane}]
Starting with the convex chain $\mathcal C$ on the sphere, we triangulate it. We denote the set of triangles obtained by $\mathcal A$, which by the definition of a convex chain together form a convex polygon (see Figure~\ref{fig:spheretoplane}). We look at $\mathcal A'$, the corresponding set of triangles in the plane, i.e, the set of triangles with same edge lengths, and with adjacency preserved. This is always achievable, as the dual of a triangulation of a convex polygon is a tree. 

We denote by $\mathcal C''$ the chain in $\mathcal A'$ corresponding to the chain $\mathcal C$ in $\mathcal A$. By Lemma~\ref{lemma:X}, we know that all the angles of the triangles are smaller, and thus all the angles on the boundary of $\mathcal A'$ are smaller. In particular $\mathcal C''$ has same edge length as $\mathcal C$, but every angle is strictly smaller; the endpoint of $\mathcal C$ and $\mathcal C''$ are at the same distance from each other. Note that $\mathcal C''$ is also a convex chain, because the convex angles of $\mathcal C$ only decreased, and therefore no reflex vertex can occur.

Recall that $\mathcal C'$ is a chain in the plane with same edge lengths and same angles as $\mathcal C$. To transform $\mathcal C''$ into $\mathcal C'$, all the angles of $\mathcal C''$ have to be increased. We apply Cauchy's Lemma in the plane, and deduce that the endpoints of $\mathcal C'$ are farther apart than those of $\mathcal C''$ and thus than those of $\mathcal C$.  
\end{proof}

\section{Cauchy's Lemma on the Growing Sphere}
In this section we generalize the result stated above to the case where we move from a sphere $\mathcal S$ to a larger sphere $\mathcal S'$. To achieve this goal, we will make use of a theorem of Legendre dating back to 1798~\cite{legendre}: 

\begin{theorem}[Legendre] 
Given a triangle $\mathcal T$ on a sphere $\mathcal S$ with angles $\theta_{1}, \theta_{2}, \theta_{3}$  and a triangle $\mathcal T'$ in the plane with same edge lengths, and with corresponding angles $\theta'_{1}, \theta'_{2}, \theta'_{3}$, we have
$$\forall i \in \{1,2,3\}, \theta'_{i}=\theta_{i} - {\delta\over 3} + (4)$$
where $\delta$ is the spherical excess of the triangle $\mathcal T$ and $(4)$ denotes a polynomial expression of the edge lengths of degree at least 4.
\end{theorem}

In other words, for sufficiently small triangles, that is, where the fourth order terms are negligible, the spherical excess is evenly split among the three angles. What the formula does not say, however, is if, for all triangles on the sphere -- even large ones -- all three angles are \emph{always} larger that their counterparts in the plane. A corollary of our results in the next section will show that this is actually the case. 

Note that Legendre's Theorem is also valid if we compare the same triangle drawn on two spheres $\mathcal S$ and $\mathcal S'$ of radius $r$ and $r'$ respectively, and with $r<r'$. In that case, its formulation is $$\forall i \in \{1,2,3\}, \theta'_{i}=\theta_{i} - {\delta-\delta'\over 3} + (4)$$ \noindent where $\delta$ and $\delta'$ are the spherical excesses on both spheres. $(\delta-\delta')$ is always positive, as long as $r<r'$.  

Here is our main Theorem:

\begin{theorem}\label{thm:sphere}
Let $\mathcal C$ be a convex chain embedded on the sphere $\mathcal S$ of radius $r$, and $\mathcal C'$ be a convex chain with same angles and edge lengths on the sphere $\mathcal S'$. 
Then the distance between the endpoints of $\mathcal C'$ is greater than the distance between the endpoints of $\mathcal C$.
\end{theorem}

To prove it, we proceed as follows, using two technical lemmas: first we show that given a thin triangle on a sphere $\mathcal S$, and a triangle with same edge lengths on a larger sphere $S'$, the small angle is strictly smaller on $\mathcal S'$. Second, we generalize that to any triangle, showing that \emph{every} angle is smaller on $\mathcal S'$. And finally, we combine these two results to prove our Theorem. 

\begin{lemma} \label{lem:thin_triangle}
Let $\mathcal T$ be a thin triangle embedded on the sphere $\mathcal S$ of radius $r$, meaning that the apex angle $\theta_{1}<\epsilon$ for some $\epsilon>0$. Let $\mathcal T'$ be a triangle with same edge lengths on the sphere $\mathcal S'$ of radius $r'>r$. Then, $\theta_{1}>\theta'_{1}$, where $\theta'_{1}$ is the angle corresponding to $\theta_{1}$ on the larger sphere.
\end{lemma}

\begin{figure}[htbp]
\centering
\includegraphics[width=0.95\linewidth]{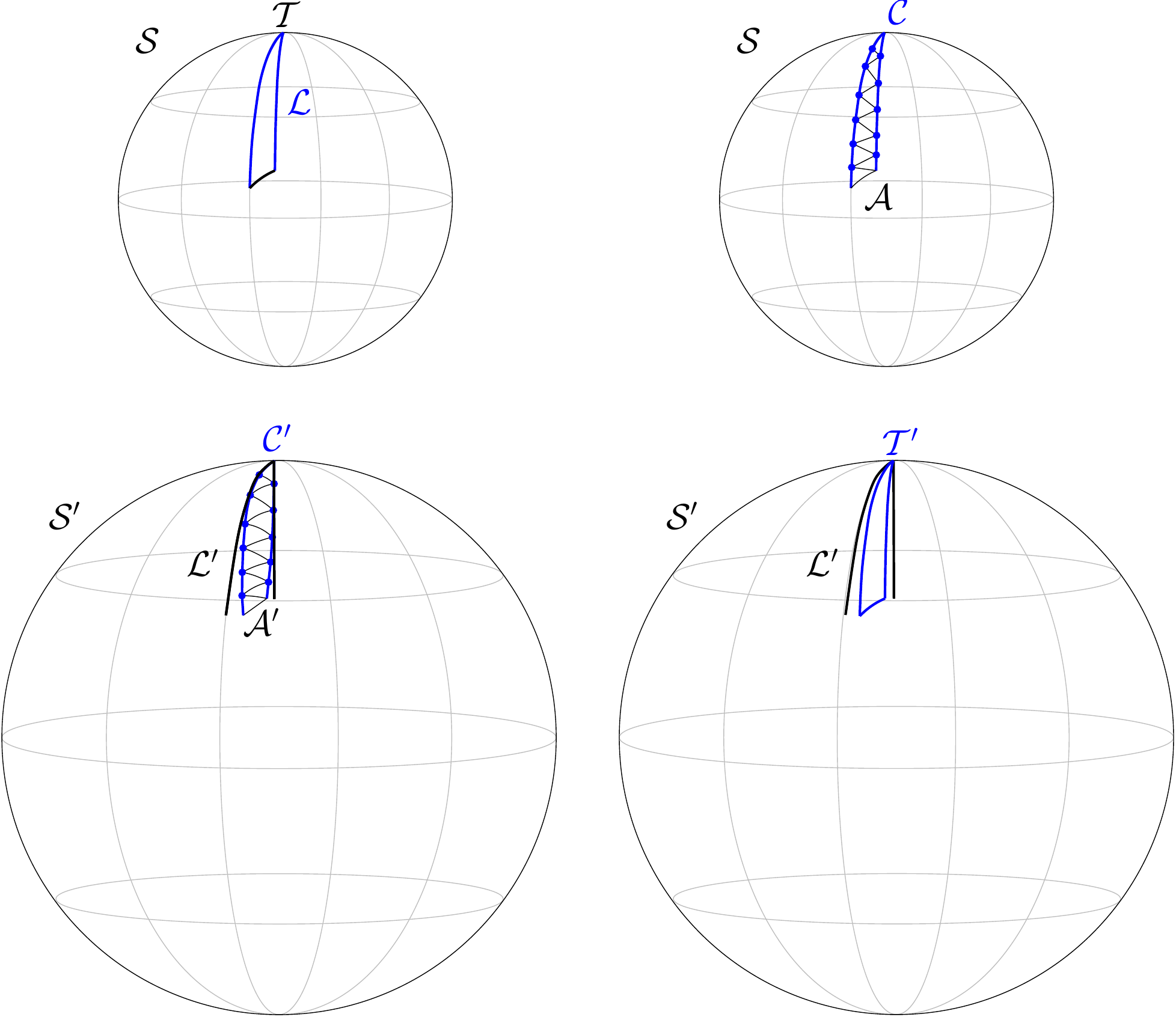}
\caption{Illustration of the proof of Lemma~\ref{lem:thin_triangle}.}
\label{fig:thintriangle}
\end{figure}

\begin{proof}
Given the radius $r$ of the sphere $\mathcal S$, the parameter $\epsilon$ defines the largest possible length of the smallest edge in $\mathcal T$, i.e., the edge opposed to $\theta_{1}$. Let $\ell = f(\epsilon, r)$ be that maximal possible length. By adding Steiner vertices on the edges of $\mathcal T$, we can triangulate it such that every triangle has edges of lengths at most $2\ell$. We denote by $\mathcal A$ the triangulation of $\mathcal T$. 

Let $\mathcal A'$ be the same set of triangles on $\mathcal S'$. As the spherical excess of every triangle on a sphere is positive, and because we can pick $\epsilon$ as small as needed -- and thus have $\ell$ and the sides of our triangles as small as needed -- the fourth order terms in Legendre's Theorem are negligible. Thus, every angle of each triangle of $\mathcal A'$ is strictly smaller than the corresponding one in $\mathcal A$. We also obtain that $\mathcal A'$ is a convex region, just as in the proof of Theorem~\ref{thm:spheretoplane}, because all the angles of the boundary of $\mathcal A$ became smaller in $\mathcal A'$. 

Now, consider the $2$-chain $\mathcal L$ composed of the two longer edges of $\mathcal T$. Let  $\mathcal C$ be the chain on the convex hull of $\mathcal A$ coinciding with $\mathcal L$, i.e., $\mathcal C$ is the subdivision of $\mathcal L$ in a chain where all edges have length less than $2\ell$. Let $\mathcal C'$ be the corresponding chain on the sphere $\mathcal S'$, defined by the convex hull of the set of triangles $\mathcal A'$. 

Starting with $\mathcal C'$, we increase all the angles to obtain a $2$-chain $\mathcal L'$, with same edge lengths and angles as in $\mathcal L$. Recall that Cauchy's Lemma is valid on a sphere, so we know that the endpoints of $\mathcal L'$ are farther apart than in $\mathcal C'$. In other words, if we want to close the $2$-chain $\mathcal L'$ to obtain the triangle $\mathcal T'$ with same edge length as in $\mathcal T$, we need to shrink the small angle. Thus, we conclude that $\theta_{1}>\theta'_{1}$, which completes the proof. 
\end{proof}

\begin{lemma}\label{lem:all_triangles}
Let $\mathcal T$ be a triangle embedded on the sphere $\mathcal S$ of radius $r$. Let $\mathcal T'$ be a triangle with same edge lengths on the sphere $\mathcal S'$ of radius $r'>r$. Then, $\theta_{i}>\theta'_{i}$ for $i\in \{1,2,3\}$, where $\theta'_{i}$ is the angle corresponding to $\theta_{i}$ on the larger sphere.
\end{lemma}

\begin{figure}[htbp]
\centering
\includegraphics[width=0.95\linewidth]{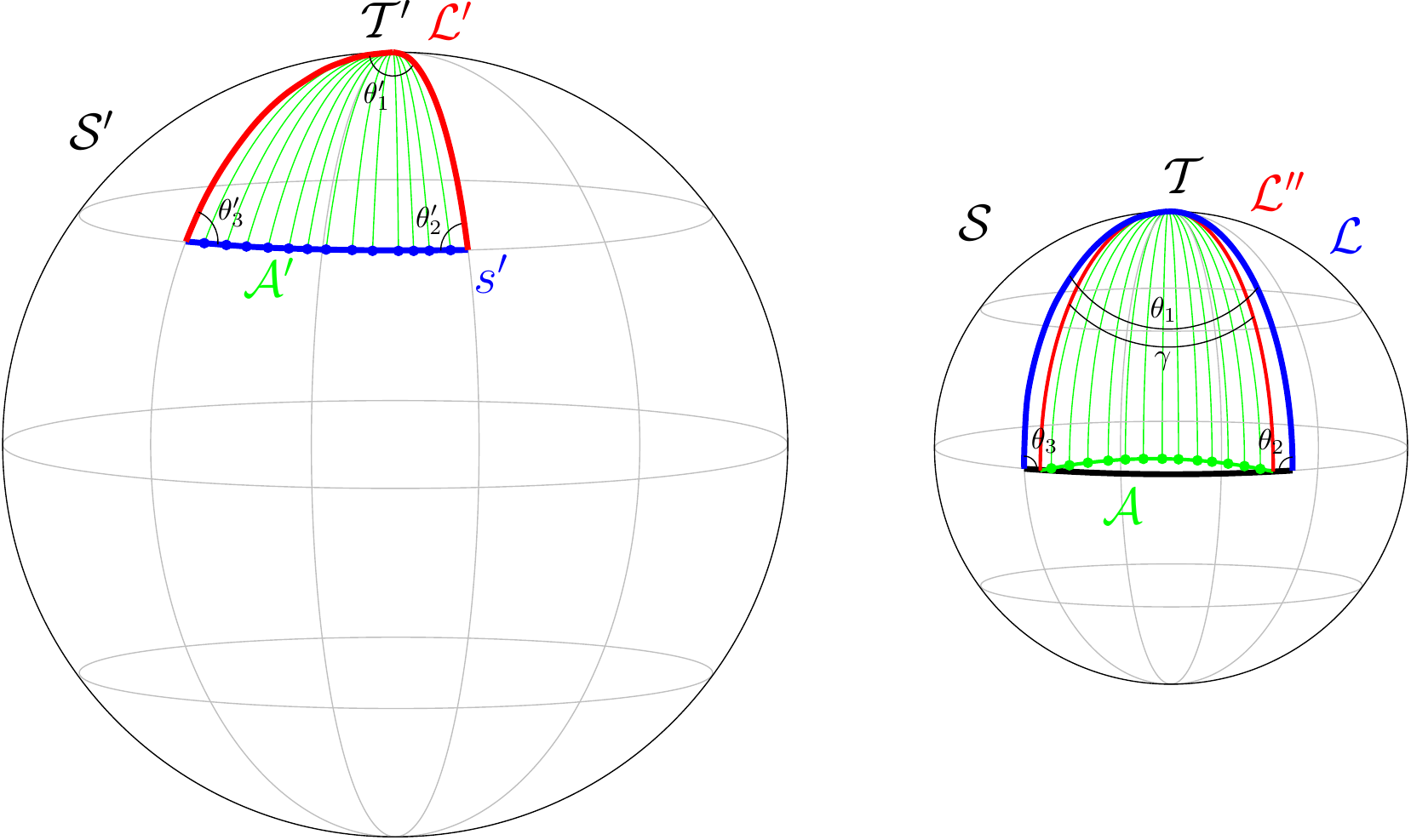}
\caption{Illustration of the proof of Lemma~\ref{lem:all_triangles}.}
\label{fig:alltriangles}
\end{figure}

\begin{proof} 
Let $s'$ be the side of $\mathcal T'$ opposed to $\theta'_{1}$, and $\mathcal L'$ denote the $2$-chain composed by the two other sides. Decompose $s'$ into as many subsegments as needed to make sure that each has length less than $\ell$, which is a value depending on $r$ and $r'$\footnote{We do not need to know $\ell$ precisely, it should just be small enough so that we can apply Lemma~\ref{lem:thin_triangle} later on.}. We denote by $\mathcal A'$ the fan-triangulation based on these subsegments. 

Let $\mathcal A$ be the corresponding set of triangles on $\mathcal S$, with edge lengths preserved. Let $\mathcal L''$ be the $2$-chain corresponding to $\mathcal L'$ in $\mathcal A$, with angle $\gamma$. Note that $s$, the chain corresponding to $s'$ on $\mathcal S$, is not straight, because by Lemma~\ref{lem:thin_triangle}, we know that the small angle of every triangle is smaller on $\mathcal S'$ than on $\mathcal S$. 
Also, we know that $\gamma$ is strictly larger than $\theta_1'$, for the same reasons. 

Because the sum of the length of the edges of $s$ is exactly equal to the length of $s'$, and because every angle in the chain is strictly less than $\pi$, the distance between the two endpoints of $\mathcal L''$ is strictly smaller than $s$. This means that the angle $\theta_{1}$ is strictly larger than $\gamma$. We conclude that $\theta_{1}'>\theta_{1}$. 

It only remains to note that we could have picked any of the three angles of $\mathcal T'$, and thus every angle is strictly larger in $\mathcal T$ than in $\mathcal T'$. 
\end{proof}

We now have all the tools required to prove Cauchy's Lemma on the growing sphere.

\begin{proof}[Proof of Theorem~\ref{thm:sphere}]
The proof is similar to that of Theorem~\ref{thm:spheretoplane}. Starting with the chain $\mathcal C$, we triangulate it, obtaining $\mathcal A$. We look at $\mathcal A'$ the corresponding set of triangles on $\mathcal S'$. By Lemma~\ref{lem:all_triangles}, we know that all the angles of the triangles are smaller, and thus all the angles on the boundary of $\mathcal A'$ are smaller. To make the boundary of $\mathcal A'$ and the chain $\mathcal C'$ coincide, we have thus to increase all angles. Using Cauchy's Lemma on the sphere (which applies by the same reasoning as in Theorem~\ref{thm:spheretoplane}), we deduce that the endpoints of $\mathcal C'$ are farther apart than those of $\mathcal C$.  
\end{proof}

In summary, when a convex chain on one sphere is redrawn (with the same lengths and angles) on a larger sphere, the distance between its endpoints increases. 

\section{Generalization to Complete Riemannian Manifolds of Positive Sectional Curvature}\label{sec:toponogov}
In this section we present a more general version of Lemma~\ref{lem:all_triangles}, and prove it by a simple application of Toponogov's Comparison Theorem~\cite{toponogov}. Here is the part of the theorem relevant to our application~\cite{toponogovsurvey}:

\begin{theorem}[Toponogov]
Let $M$ be a complete Riemannian manifold\footnote{A \emph{Riemannian manifold} is a differentiable manifold whose tangent spaces support a smoothy varying inner product, which permits angles and lengths to be defined. It is \emph{complete} if Cauchy sequences have limits.} with sectional curvature\footnote{The \emph{sectional curvature} measures the deviation of geodesics, and is the natural generalization of Gaussian curvature for surfaces. A sphere has constant positive sectional curvature.} $K\ge \kappa$.
Given points $p_{0},p_{1},q$ in $M$ satisfying $p_{0}\neq q , p_{1}\neq q$, a non constant geodesic\footnote{A \emph{non constant geodesic} is any geodesic of positive length.} $c$ from $p_{0}$ to $p_{1}$ and minimal geodesics $c_{i}$, from $p_{i}$ to $q$, $i=0,1$, all parameterized by arc length. Suppose the triangle inequality $|c|\le|c_{1}| + |c_{2}|$ is satisfied and $|c|\le{\pi\over \sqrt{\kappa}}$ in the case $\kappa\ge 0$. $\alpha_i\in[0,\pi]$ denote the angles at $p_{i}, \alpha_{0}= \angle(\dot{c}_{0}(0),\dot{c}(0))$, $\alpha_{1}=\angle(\dot{c}_{1}(0),-\dot{c}(|c|))$. Then there exists a corresponding comparison triangle $\tilde{p_{0}}, \tilde{p_{1}} , \tilde{q}$ in the model space $M^2_{\kappa}$ with corresponding geodesics $\tilde{c_{0}},\tilde{c_{1}},\tilde{c}$ which are all minimal of length $|\tilde{c_{i}}|=|c_{i}|, |\tilde{c}|=|c|$ and the corresponding angles $\tilde{\alpha_{i}}$ satisfy $\tilde{\alpha_{i}}\le \alpha_{i}$.
\end{theorem}

The interesting part for the application we have in mind is to notice that the two spheres we consider in the previous sections are actually model spaces $M^2_{\kappa}$, with $\kappa$ depending on the radius of the sphere. So, we obtain directly a generalized version of the lemma used in previous sections:

\begin{corollary}\label{lem:all_triangles_toponogov}
Let $\mathcal T$ be a triangle embedded on a complete Riemannian manifold $M_{\kappa}$ with sectional curvature $\kappa$ and where the triangle inequality holds. Let $\mathcal T'$ be a triangle with same edge lengths on a model space $M^2_{\kappa'}$ with sectional curvature $\kappa'<\kappa$, also satisfying the triangle inequality. Then, $\theta_{i}>\theta'_{i}$ for $i\in \{1,2,3\}$, where $\theta'_{i}$ is the angle corresponding to $\theta_{i}$ on $\mathcal R'$.
\end{corollary}

\noindent We can use this to get an alternate proof of Theorem~\ref{thm:sphere}, with the same arguments, using the model spaces $M^2_{\kappa}$ and $M^2_{\kappa'}$ which are two spheres of radius $r={1\over\sqrt{\kappa}}$ and $r'={1\over\sqrt{\kappa'}}$. \bigskip

\begin{openproblem}What is the class of surfaces on which the original version of Cauchy's Arm Lemma (Lemma~\ref{lem:cauchy}) is true?
\end{openproblem}
The answer to this question would provide the class of surfaces on which we could generalize Cauchy's Lemma, i.e., redrawing the same chain (angles and lengths preserved) on two different surfaces and deducting that the endpoints move farther apart. We can reuse the same reasoning as long as (the original version of) Cauchy's Arm Lemma and Toponogov's Theorem apply to the surfaces we consider. 

\section*{Acknowledgments}

This note was prepared in part during the \emph{23$^{rd}$ Bellairs Winter Workshop
on Computational Geometry} held February 1--8, 2008, and organized by Godfried Toussaint.
We thank the participants of that workshop, particularly Brad Ballinger,
for helpful discussions.

{
\bibliographystyle{plain}
\bibliography{note}

\begin{thebibliography}{1}

\bibitem{Cromwell}
P.~Cromwell.
\newblock {\em Polyhedra}.
\newblock Cambridge University Press, 1997.

\bibitem{legendre}
A.M. Legendre.
\newblock Méthode pour déterminer la longueur exacte du quart du méridien
  d'après les observations faites pour la mesure de l'arc compris entre
  dunkerque et barcelone, note iii: Résolution des triangles sphériques dont
  des côtés sont très petits par rapport au rayon de la sphère.
\newblock {\em Méthodes analytiques pour la détermination d'tun arc du
  méridien}, pages 12--14, 1798.

\bibitem{toponogovsurvey}
W.~Meyer.
\newblock {\em Toponogov's Theorem and Applications}.
\newblock College on Differential Geometry at Trieste, 1989.

\bibitem{toponogov}
V.A. Toponogov.
\newblock Riemannian spaces having their curvature bounded below by a positive
  number.
\newblock {\em Amer. Math. Soc. Transl. Serv.}, 37:291--336, 1964.

\end{thebibliography}
}
\end{document}